\documentclass[aps,prd,twocolumn,
showpacs,
floatfix,nofootinbib,
superscriptaddress]{revtex4}

\usepackage{pstricks,epsfig,color}
\usepackage{psfrag}

\newcommand{\be}{\begin{equation}}
\newcommand{\ee}{\end{equation}}
\newcommand{\bear}{\begin{eqnarray}}
\newcommand{\eear}{\end{eqnarray}} \newcommand{\ba}{\begin{array}}
\newcommand{\ea}{\end{array}}
\newcommand{\lae}{\begin{array}{c}\,\sim\vspace{-1.7em}\\< 
\end{array}}

\def\beq{\begin{equation}}
\def\eeq#1{\label{#1}\end{equation}}
\def\eeqn{\end{equation}}
\def\eeq{\end{equation}}
\def\beqa{\begin{eqnarray}}
\def\eeqa#1{\label{#1}\end{eqnarray}}
\def\eeqan{\end{eqnarray}}

\def\to{\rightarrow}

\newcommand\iden{\leavevmode\hbox{\small1\normalsize\kern-.33em1}}

\def\W3{W_H^3}

\begin{document}

\title{New Strongly Coupled Sector at the Tevatron and the LHC}
\author{Gustavo Burdman
}
\author{
Leonardo de Lima
} 
\author{Ricardo D. Matheus
\footnote{Current address: {\em Departamento de Ciências Exatas e da Terra, Universidade Federal de
São Paulo, Diadema SP 09972-270, Brazil}}
}
\affiliation{Instituto de F\'isica, Universidade de S\~ao Paulo, 
S\~ao Paulo SP 05508-900, Brazil}
\vspace*{0.3cm}


\begin{abstract}
We examine the possibility that a new strong interaction is 
accessible to the Tevatron and the LHC.
In an effective theory approach, we consider a scenario with a new
color-octet interaction with strong couplings to the top quark, as well as the
presence of a strongly coupled fourth-generation which could be
responsible for electroweak symmetry breaking. We apply several
constraints, including the ones from flavor physics. 
We study the phenomenology of the resulting parameter space at the Tevatron,
 focusing on the  the forward-backward asymmetry in
top pair production, as well as in the 
production of the fourth-generation
quarks. We show that if the excess in the top production asymmetry is indeed the
result of this new interaction, the Tevatron could see the first hints
of the strongly coupled fourth-generation quarks. Finally, we show
that the LHC with $\sqrt{s}=7~$TeV and $1~{\rm fb}^{-1}$ integrated
luminosity should observe the production of fourth-generation quarks
at a level at least one order of magnitude above the QCD prediction
for the production of these states. 
\end{abstract}

\maketitle

\section{Introduction}
\label{intro}
Although the Standard Model (SM) of particle physics is a very successful
description of the interactions of fermions and gauge bosons~\cite{pdg}, the
origin of electroweak symmetry breaking remains unknown. In the SM, an
ad hoc scalar sector with an elementary doublet is responsible for triggering 
spontaneous symmetry breaking, giving masses to gauge bosons and
fermions. However, there are several reasons to believe that this
description is likely to be an effective one. 
First, the weak scale as generated by the elementary scalar Higgs
sector is not stable under radiative corrections, resulting in an
unnatural tuning. Perhaps most importantly, it appears more natural
to have the scalar sector as a composite of a fermionic sector, given
our experience with other physical systems. The only known loophole to this
statement is found in supersymmetric theories, where elementary
scalars are natural and the weak scale can be stabilized by the
presence of super-partners. Here we assume that supersymmetry is not 
present at the weak scale and therefore the Higgs sector must be a
fermionic composite, with the compositeness scale not far above the weak scale.
In particular, we assume that the new strong interactions are spontaneously 
broken, and therefore the condensing fermions are not confined
by it. This opens the intriguing possibility that 
the condensing strongly coupled fermions might belong to a sequential 
fourth-generation. This scenario~\cite{bhl} differs from technicolor theories~\cite{tc}
where techni-fermions are confined by an unbroken, asymptotically free 
interaction. Although it is relatively simple to build a renormalizable model
along the lines of top-color~\cite{topc,fgewsb} in order to obtain the condensation
of a fourth-generation quark, a more complete model (e.g. including 
mass generation for  all fermions) is more elusive. 
For instance, recently 
a model embedded in AdS$_5$ was presented in Ref.~\cite{bd,bdem}, where 
the fourth-generation is strongly coupled to the Kaluza-Klein (KK) 
excitations of the gauge bosons due to its localization  in a
compact extra dimension.

In this paper we would like to focus on  the basic ingredients for  this scenario: a new strong interaction at the
TeV, a fourth-generation strongly coupled to it, and with enough
flavor violation to generate the flavor hierarchies. 
The aim is to apply the minimal set of requirements to a model of 
fourth-generation condensation in order to fix some important aspects of its 
phenomenology at colliders.  One important feature that must be
present is flavor violation at tree level, ensuring a super-critical
coupling of the fourth-generation quarks. It is then natural to assume
that he third generation might also be more strongly coupled to the
mass-generating interaction than the lighter two.  Thus, we are
inclined not just to consider the fourth-generation phenomenology in
isolation, but also the signals and constraints from flavor violation
involving the third generation. In particular, we study the
possibility that the new interaction coupled to the top quark might
result in large deviations in the top-production forward-backward
asymmetry $A_{FB}^t$ recently measured at the Tevatron.  At the same
time, we must consider the flavor physics bounds on flavor-violating processes.

Rather than to attempt building a full fledged theory we will take an
effective field theory approach and write the most general
interactions containing these ingredients and satisfying all existing
constraints. This procedure will be restrictive enough so as to
result in a predictive model of the  new interaction, including the
fourth generation quarks. 
 In order to implement this approach,  
we consider a full fourth generation of chiral fermions to have an 
anomaly-free extension from the start.
 We require a new interaction strongly 
coupled to at least some of the fourth-generation fermions.
However, in this paper we will not be concerned with the phenomenology
of the fourth-generation leptons~\cite{fgleptons}, since early signals
of this scenario are much more likely in the quark sector.
 
We further assume 
that the new interaction is spontaneously broken at a scale close to 
1 TeV, and that it is mostly mediated by a color-octet spin-one massive state.
Although this choice is not unique, it does appear in various models such as
the fourth-generation version of top-color and extra dimensional theories.
Finally, we demand that the interaction couples to the 
fourth-generation quarks strongly enough so as to induce $\langle Q_4U_4\rangle\not=0$ 
and/or $\langle Q_4D_4\rangle\not=0$, where $Q_4$, $U_4$ and $D_4$ are
the  doublet, up right-handed and down right-handed fourth-generation quarks
respectively. If at least one of these two condensates forms, it
induces electroweak symmetry breaking 
(EWSB) and generates a dynamical mass for the condensing quark. 
The requirement of super-critical coupling of the color-octet to the 
fourth-generation quarks is important because it greatly determines its
width. In fact, as we will see later, the width of the color-octet must be
rather large in all of these scenarios. 

Regarding the lighter three 
generations, their masses will typically arise from higher dimensional 
operators, which are too suppressed to affect the phenomenology at colliders.
On the other hand, the color-octet couplings to the SM quarks should
be considerably smaller than the couplings to the fourth generation.
Here we will mostly leave these couplings free, but for the 
constraints imposed on them by flavor-changing neutral currents (FCNC),  by multijet 
production and by top quark observables. As a result, we 
will obtain the allowed parameter space for this scenario which will result
in predictions for the production of fourth-generation quarks via this 
new interaction plus QCD both at the Tevatron and at the LHC.
We first focus our attention on the potential of the Tevatron to 
produce the fourth-generation quarks given that it will eventually 
accumulate $10^{-1}fb$ per experiment. In addition, we will consider the
possibility that the new color-octet interaction, if appropriately coupled to 
top-quarks, could be responsible for the observed deviation 
in the measured forward-backward asymmetry in top-quark production at the 
Tevatron~\cite{cdfafb} with respect to the SM prediction. Once this
additional information from the Tevatron is considered, we study the
LHC reach during its first physics run, with $\sqrt{s}=7$ ~TeV and
$1^{-1}fb$ of accumulated luminosity.

There is a large number of existing studies involving either a new interaction leading to deviations
in $A_{FB}^t$~\cite{refsafbtop}  or the phenomenology of the  fourth generation~\cite{refsfgen}. This is
the first attempt to combine the two in one effective
model. Particularly relevant to our study on $A_{FB}^t$ is
Ref.~\cite{frampton}, where an axi-gluon model is studied. 
In Ref.~\cite{sekhar} this specific  axi-gluon model is excluded by
flavor constraints.    

In the next section we define the effective theory to be used in the rest 
of the paper. In Section~\ref{parspace} we fix the parameter space of the 
model by requiring it to satisfy FCNC constraints, as well as various 
direct detection observables including the observed $A_{FB}$ in top-quark 
production. In Section~\ref{fg}, the fourth-generation quarks are
introduced and the constraints on them are presented. 
In Section~\ref{predictions} we present our results and evaluate the
reach of the Tevatron in both the color-octet and the
fourth-generation masses. We also discuss the level of these signals
at the LHC. We conclude in Section~\ref{conclusions}.

\section{Effective  Theory}
\label{efftheory} 
We extend the SM by including a chiral fourth generation 
$Q_4, U_4, D_4, L_4, E_4, N_4$. We also assume the presence 
of a massive, color-octet, spin-one state $G^a_\mu$. The relevant
effective interaction with quarks is given by
\bear
{\cal L}_{\rm eff} &=& g_L^i\,G_\mu^a \,\bar Q_i\gamma^\mu T^a Q_i
\nonumber\\
&+& g^i_u \,G_\mu^a \,\bar U_i\gamma^\mu T^a U_i \nonumber\\
&+& g^i_d \,G_\mu^a \,\bar D_i\gamma^\mu T^a D_i~,
\label{leff}
\eear 
where $T^a$ are the $SU(3)_c$ generators, 
a sum over the generation number $i=1,2,3$ is understood,  $Q_i$
denotes the left-handed quark doublet and  
$U_i (D_i)$ the up (down) right-handed quark of the i-th generation.  
Although at this point the couplings $g^i_L$, $g^i_u$ and $g^i_d$ are
free parameters, we will impose constraints on them, some of which come
from the desired dynamics of EWSB, whereas other will be purely
phenomenological in origin.

In order to focus in a scenario where EWSB is triggered by the
condensation of fourth-generation quarks we ask that the
fourth-generation couplings be strong enough to lead to at least one
of the two quark condensates,  $\langle Q_4U_4\rangle$ and 
$\langle Q_4D_4\rangle$, to be non-vanishing. The four-fermion 
operators of interest are induced by integrating out the color-octet
and are given by 
\bear
{\cal L}_{\rm eff}^4 &=& \frac{g_L^4 g^4_u}{M_G^2}\, \bar Q_4\gamma_\mu
T^a U_4 \,\bar U_4\gamma^\mu T^a Q_4 \nonumber\\
&+&   \frac{g_L^4 g^4_d}{M_G^2} \,\bar Q_4\gamma_\mu
T^a D_4 \,\bar D_4 \gamma^\mu T^aQ_4~.
\label{fgffermion}
\eear 
In order for one of these two terms to lead to condensation, 
at least one of the criticality conditions must be satisfied. That is 
\be
g_L^4 g^4_u > \frac{8 \pi^2}{3},\qquad {\rm and/or}\qquad 
g_L^4 g^4_d > \frac{8 \pi^2}{3}~.
\label{critcond}
\ee
Thus, the scenario where at least one of the fourth-generation quarks
condenses requires that the left-handed and/or the right-handed quarks
be strongly coupled to the color-octet interaction. Although these
couplings are required to be close  to non-perturbativity,
it is possible to have condensation with couplings satisfying
$g^4_{L,u,d}\lae 2\pi$, which we take as an upper limit to
the couplings we will consider here.  

On the other hand, the values of $g_L^i$, $g_u^i$ and $g_d^i$ for
$i=1,2,3$ can  be generically  smaller than the corresponding
couplings for the fourth generation. We will essentially consider two
constraints on these couplings. First, we require that for a given
color-octet mass, the $G'$ couplings of light quarks satisfy bounds
from searches for di-jet resonances at hadron colliders~\cite{dijet}.   
The bounds, for fixed values of $M_G$,  translate on limits on
$g_L^q$ and $g_R^q$ for the light quarks~\cite{bogdan}.  Typically, for $M_G\simeq 1$~TeV the
light-quark couplings to $G'$ are bound to be smaller than the QCD
coupling, and they can be almost as large as this one (i.e. $O(1)$) for $M_G\simeq
1.5~$TeV.

A second requirement on the $G'$ couplings to the first three
generation quarks is that they do not violate flavor bounds. 
In principle, the color-octet interactions violate flavor at tree level
since they couple to the fourth generation quarks with a larger
strength. However, in order to evade strong FCNC bounds from flavor
physics we will assume that the $G'$ couplings of the first three
generations are nearly universal, with the only exception of the
couplings to $t_R$. As we will see later, this is the minimum flavor violation required in
order to accommodate  a significant forward-backward asymmetry in
$t\bar t$ production at the Tevatron. 
 This flavor violation leads to significant contributions to vertices
 involving the fourth and third generation quarks, such as $G Q_4
 t_R$, but it does not generate dangerous FCNC contributions  in low
 energy flavor observables. Thus, these two requirements define an
 effective theory of a strongly coupled fourth generation, where the
 new interactions are at the TeV scale, but without problems with  flavor
 violation. 

The effective theory defined above should encompass models of
EWSB  via fourth-generation condensation that
somehow manage to avoid having large tree-level FCNC.  In the next section we
investigate the parameter space of these theories. In particular, we
consider the possibility of large contributions to the $t\bar t $
forward-backward asymmetry $A_{FB}^t$, as well as the potential of the Tevatron
to observe the production of fourth-generation quarks.

\section{Fixing the Parameter Space}
\label{parspace}
\begin{figure}
\includegraphics[scale=0.60]{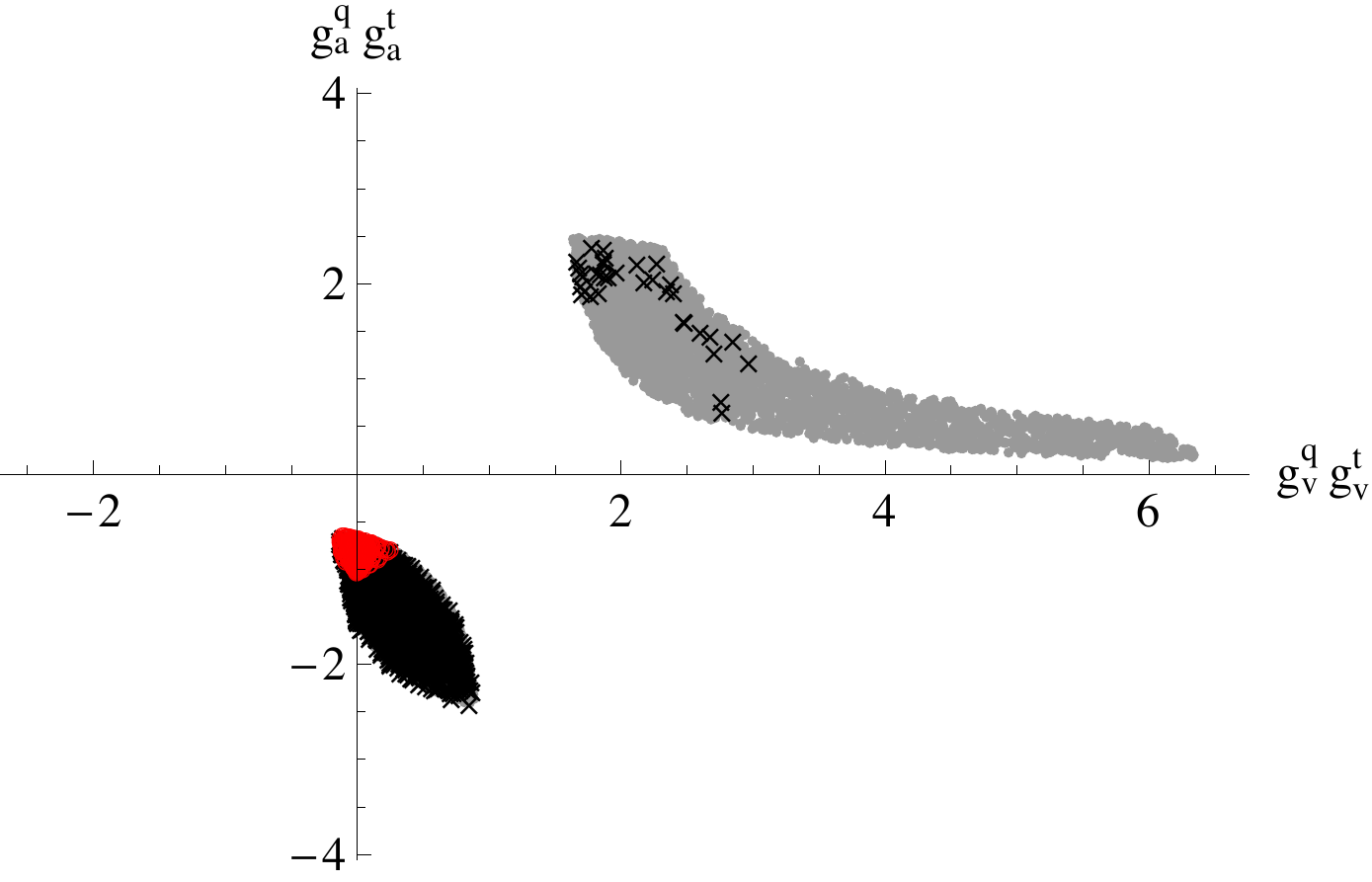}
\caption{Allowed region of parameter space leading to significant effects in
  $A_{FB}^t$, for $M_G=1$~TeV (grey). The black crosses mark the
  region of parameter space further satisfying the dijet bounds as
  well as flavor universality.
The (red) dots satisfy, in addition, the bounds from  the
invariant mass distribution of Figure~\ref{mttdist} and the
$\Delta y_t$  distribution of $A_{FB}^t$ shown in Figure~\ref{afbvsdy} . } 
\label{scat1}
\end{figure}

\begin{figure}
\includegraphics[scale=0.60]{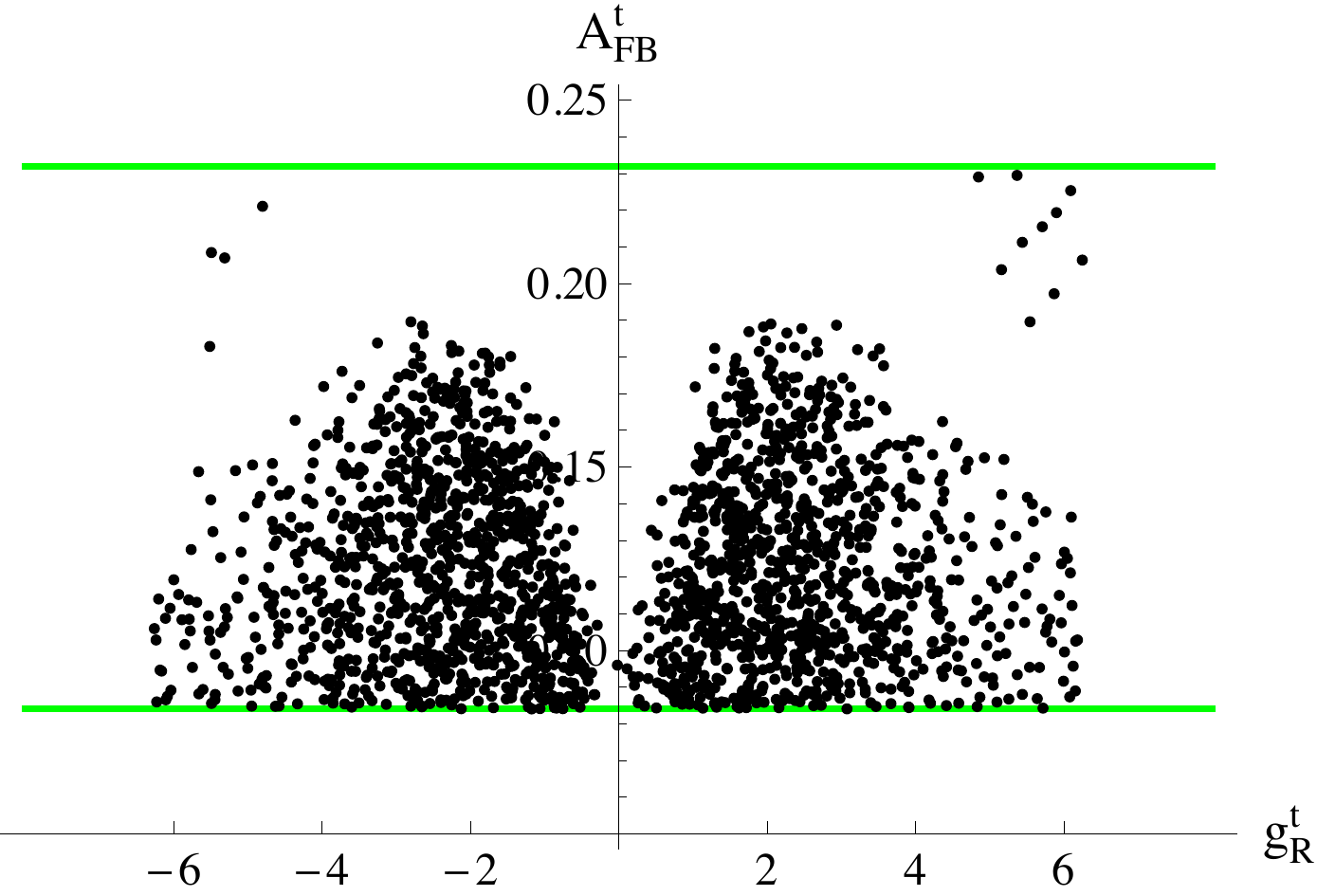}
\caption{$A_{FB}^t$ vs. the right-handed top quark coupling $g_R^t$,
  for $M_G=1$~TeV. } 
\label{afbgtr1}
\end{figure}

\begin{figure}
\includegraphics[scale=0.60]{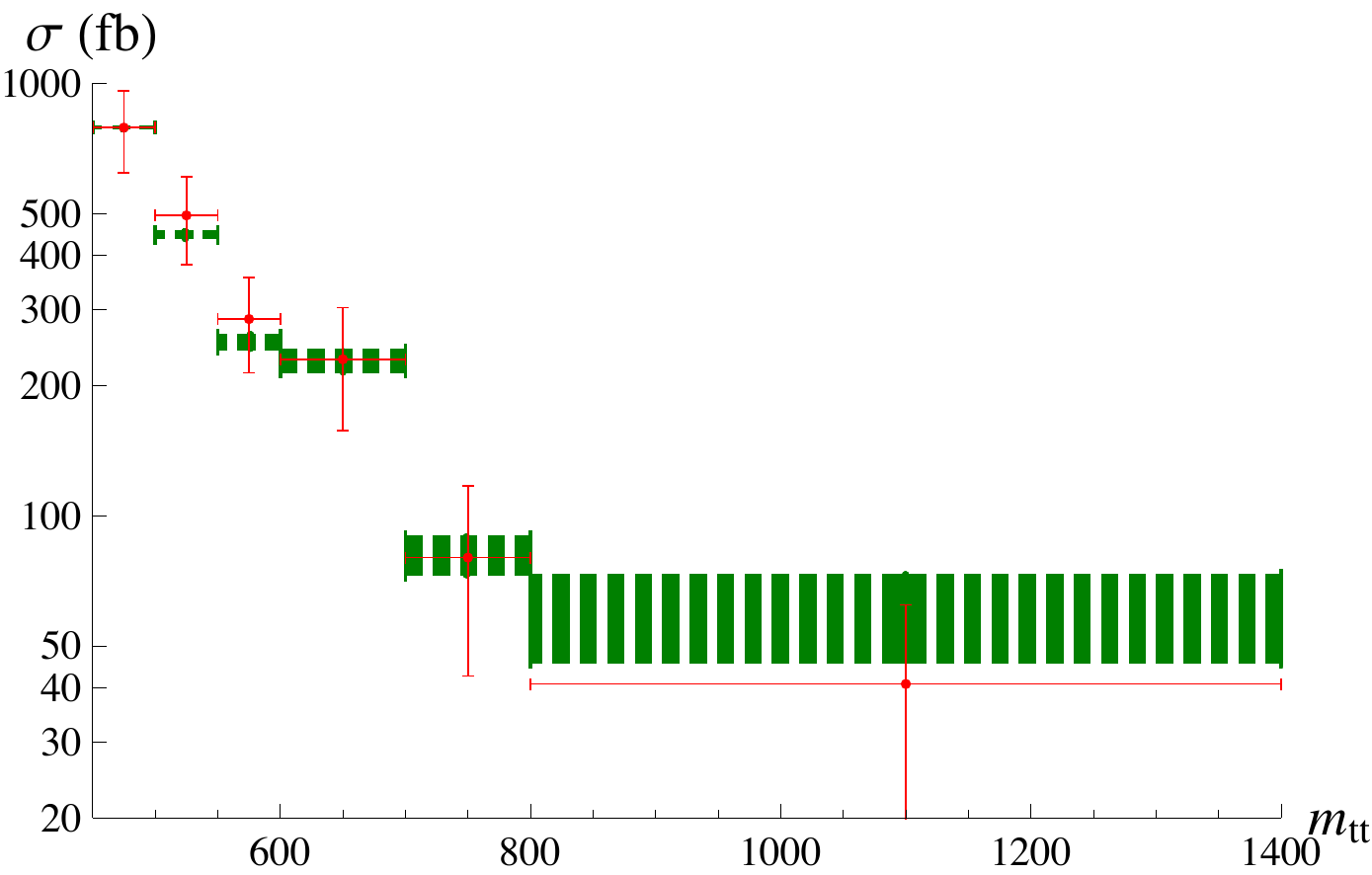}
\caption{The $t\bar t$ production cross-section as a function of the 
$t\bar t$ invariant mass $m_{tt}$, for $M_G=1$~TeV.  The crosses are
the experimental values. The band represent the solutions of the
scattered plot in Figure~\ref{scat1}  that are within 1.5 $\sigma$ of
the data. 
}
\label{mttdist}
\end{figure}

\begin{figure}
\includegraphics[scale=0.60]{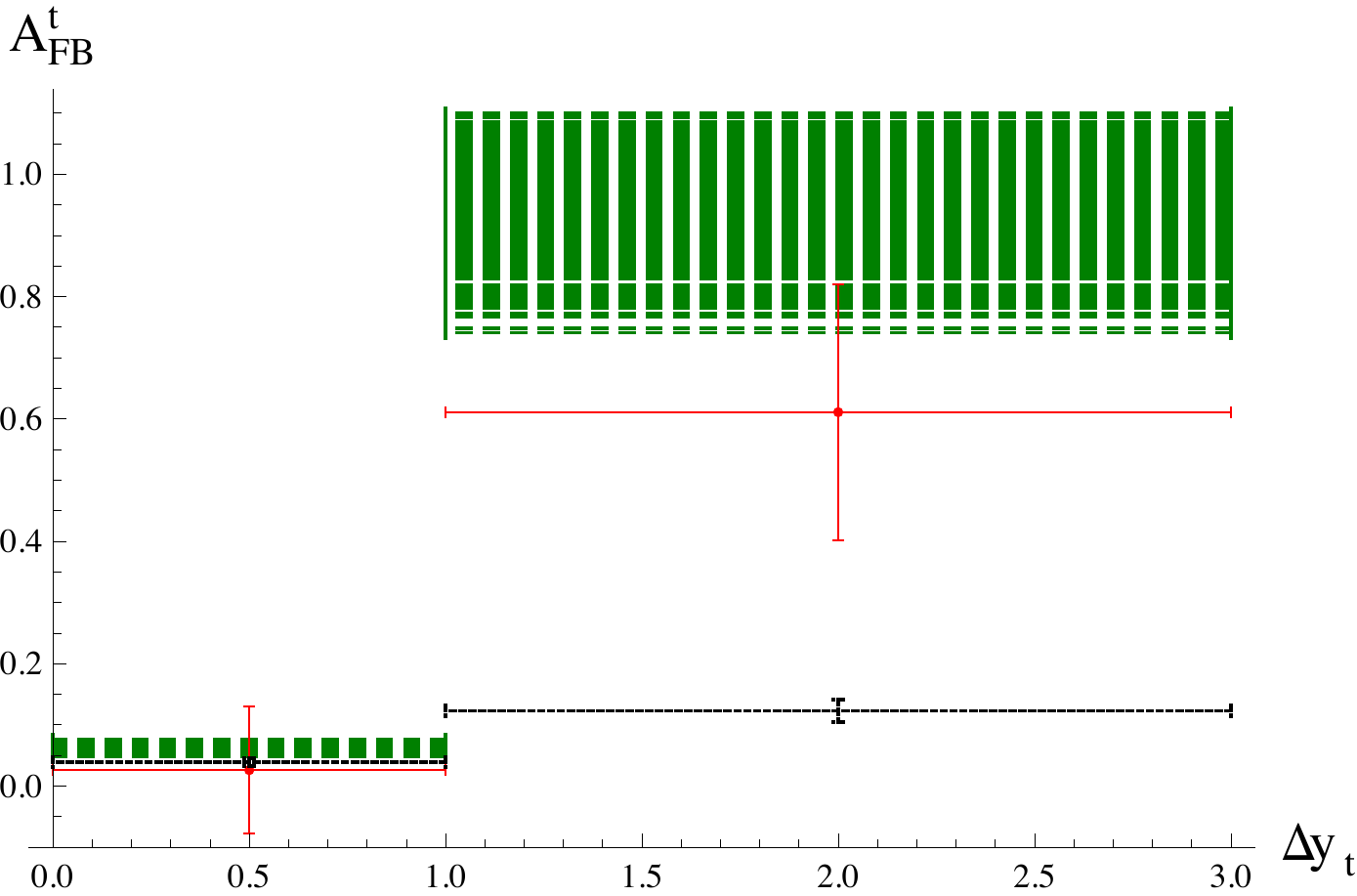}
\caption{The rapidity dependence of the forward-backward asymmetry in
  $t\bar t$ production. for $M_G=1$~TeV.  
 The crosses are
the experimental values. The band represent the solutions of the
scattered plot in Figure~\ref{scat1}  that are within 1.5 $\sigma$ of
the data.  The dashed line is the NLO QCD prediction obtained using
MCFM~\cite{mcfm}.  
}
\label{afbvsdy}
\end{figure}

In this section we reduce the size of the parameter space of the effective
theory presented in the previous section by requiring that it gives a
significant contribution to $A_{FB}^t$, while not violating FCNC
bounds. The current measurement of $A_{FB}^t$  from
the CDF collaboration at Fermilab gives~\cite{cdfafb}
\beq
A_{FB}^t = 0.158 \pm 0.072\pm  0.017~.
\label{afbtop}
\eeq
where we consider the forward-backward asymmetry in the $t\bar t $
rest frame. 
The SM prediction from NLO QCD using MCFM~\cite{mcfm}  results in  $A_{FB}^{SM} =
0.058\pm0.009$, leaving then considerable room for potential contributions
from new physics.  First, for a given value of $M_G$, we require that $g_L^q$, $g_R^q$, $g_R^t$ and
$g_L^t$ be such that the contributions of the color-octet to $t\bar t$
production result in $A_{FB}^{\rm new} = 0.16\pm 0.07$.
Figure~\ref{scat1} shows the response of the parameter space of the
effective theory to various constraints for
$M_G=1~$TeV. The shaded region of the plot shows the  selected parameter space in terms
of the product of the light quark and top  vector couplings, $g_V^q
g_V^t$, and the product of their axial couplings $g_A^q g_A^t$.  
This region results from imposing the $A_{FB}^t$ 
 constraint, plus demanding that 
 the total $t\bar t$ cross section be
within one sigma of~\cite{topxs} 
$\sigma_{t\bar t} = 7.50\pm
0.31\pm0.34\pm0.15$.
Imposing 
the perturbativity of the top quark coupling $g^t_R$,  the dijet
bounds, and flavor universality in all couplings with the exception of $g_R^t$
results in the  smaller regions
represented by the darker crosses in the Figure.
Imposing the light flavor universality leaves us with the right-handed top as the only free
coupling to achieve a large asymmetry. This is shown in
Figure~\ref{afbgtr1}, where we show solutions for $A_{FB}^t$ 
for given values of $g_R^t$ that satisfy all other constraints, and
for $M_G=1~$TeV. We see that it is possible to generate important
contributions to $A_{FB}^t$ even for moderate values of $g_R^t$. 
Furthermore, we require a good fit to the measured~\cite{mttexp}
$t\bar t$ invariant mass distribution, by excluding solutions which
would make any one bin in the distribution fall outside a $1.5~\sigma$
band. We show the result for the 
invariant mass distribution in $t\bar t$ production in
Figure~\ref{mttdist}. 
 Finally, the resulting parameter space is used to plot the
rapidity  dependence of $A_{FB}^t$ as obtained in
\cite{cdfafb},   resulting in the bands
of Figure~\ref{afbvsdy}.  
In both cases, the crosses represent the data points. We see that both distributions can be safely
accommodated with the available parameter space. In the case of the
rapidity distribution of Figure~\ref{afbvsdy}, the band representing
the available solutions is consistent with the data even when the QCD
(shown as the dashed line ) is not.  
The region of parameter space consistent with the $m_{tt}$ distribution
and with the $\Delta y$ dependence of $A_{FB}^t$, is shown in
Figure~\ref{scat1} as the smaller (red) region in the lower left
region, corresponding to low values of the product of the vector
couplings $g_v^q g_v^t$. These solutions then, satisfy all constraints.
We observe that
the selected region of parameter space is  ``axi-gluon like'', in that
the product of the vector couplings of the light quarks and the top is
small (is zero in axi-gluon models). It is also consistent with the
choice of signs of Ref.\cite{frampton}, but here the solutions are
more generic and are arrived at by imposing all constraints.

\begin{figure}
\includegraphics[scale=0.60]{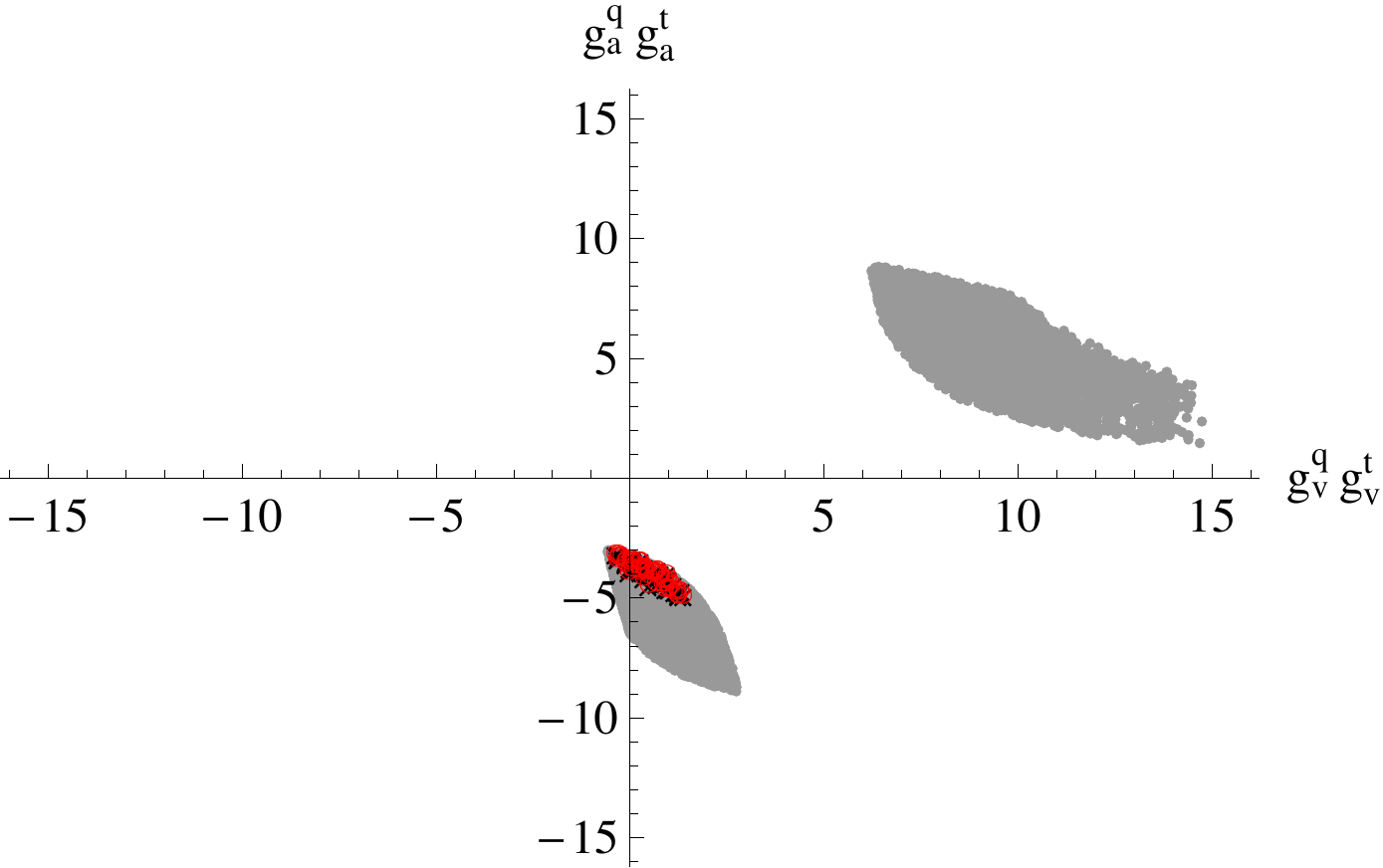}
\caption{Allowed region of parameter space leading to significant effects in
  $A_{FB}^t$, for $M_G=1.5$~TeV (grey). The black crosses mark the
  region of parameter space further satisfying the dijet bounds as
  well as flavor universality.
The (red) dots satisfy, in addition, the bounds from  the
invariant mass distribution of Figure~\ref{mttdist_15} and the
$\Delta y_t$  distribution of $A_{FB}^t$ shown in Figure~\ref{afbvsdy_15}. } 
\label{scat15}
\end{figure}

\begin{figure}
\includegraphics[scale=0.60]{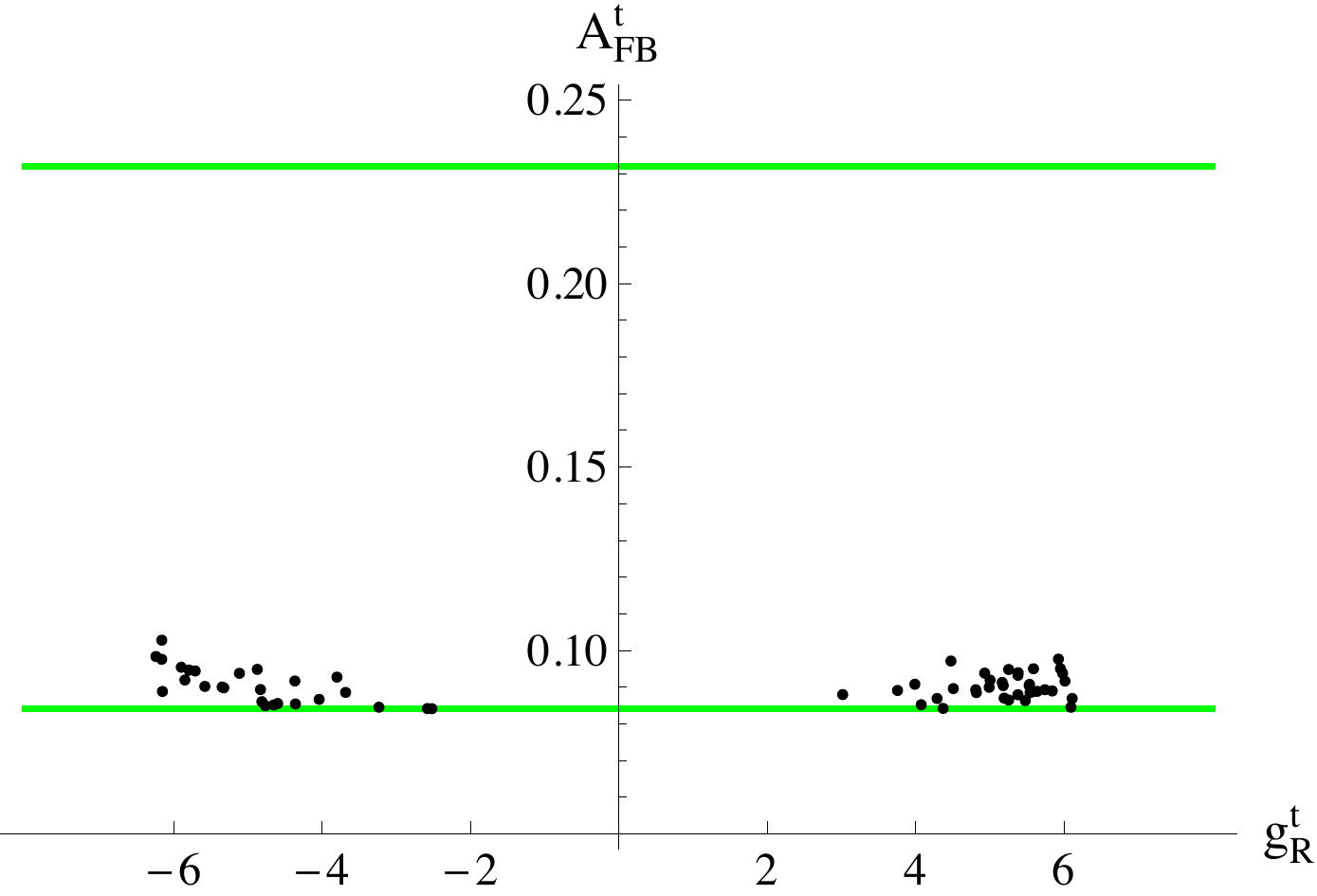}
\caption{$A_{FB}^t$ vs. the right-handed top quark coupling $g_R^t$,
  for $M_G=1.5$~TeV. } 
\label{afbgtr15}
\end{figure}

We repeat the exercise for $M_G=1.5~$TeV and show the allowed regions
in Figure~\ref{scat15}. 
Although, in principle, the region of allowed parameter space appears to be slightly larger 
for $M_G=1.5$~TeV without including the flavor conservation
constraints, we see that once these are considered only one of the two
regions  is still allowed. This is shown as (red) dots in the
lower region of Figure~\ref{scat15}. In Figure~\ref{afbgtr15}  we plot
the asymmetry vs. the right-handed top quark
coupling to $G'$. 
Just as for the previous case, in order to get a
significant effect in $A_{FB}^t$, the  
right-handed top quark must have a rather large coupling. typically
$|g_R^t|\sim (2-5)$. Although the larger values of $g_R^t$ are close
to the upper bound given by perturbativity, this is not a problem
since in order for the top quark to condense the left-handed coupling
to $G'$ should be of the same order. But this is not allowed by the
flavor-conservation constraint.
In addition,  we have to allow for larger values of
the light-quark couplings $g_L^q$ and $g_R^q$, although for this value
of $M_G$ these remain typically just below the QCD coupling.
In principle, it is possible to consider larger values of the $G$ mass. 
To go above $M_G=1.5~$TeV without significantly increasing  the
value of $g_R^t$ --already at the edge of perturbativity--
would require light-quark couplings above the QCD coupling. But in doing
so we would run into trouble with the bounds on resonances decaying to
two jets from the Tevatron data~\cite{dijet}. 
\begin{figure}
\includegraphics[scale=0.60]{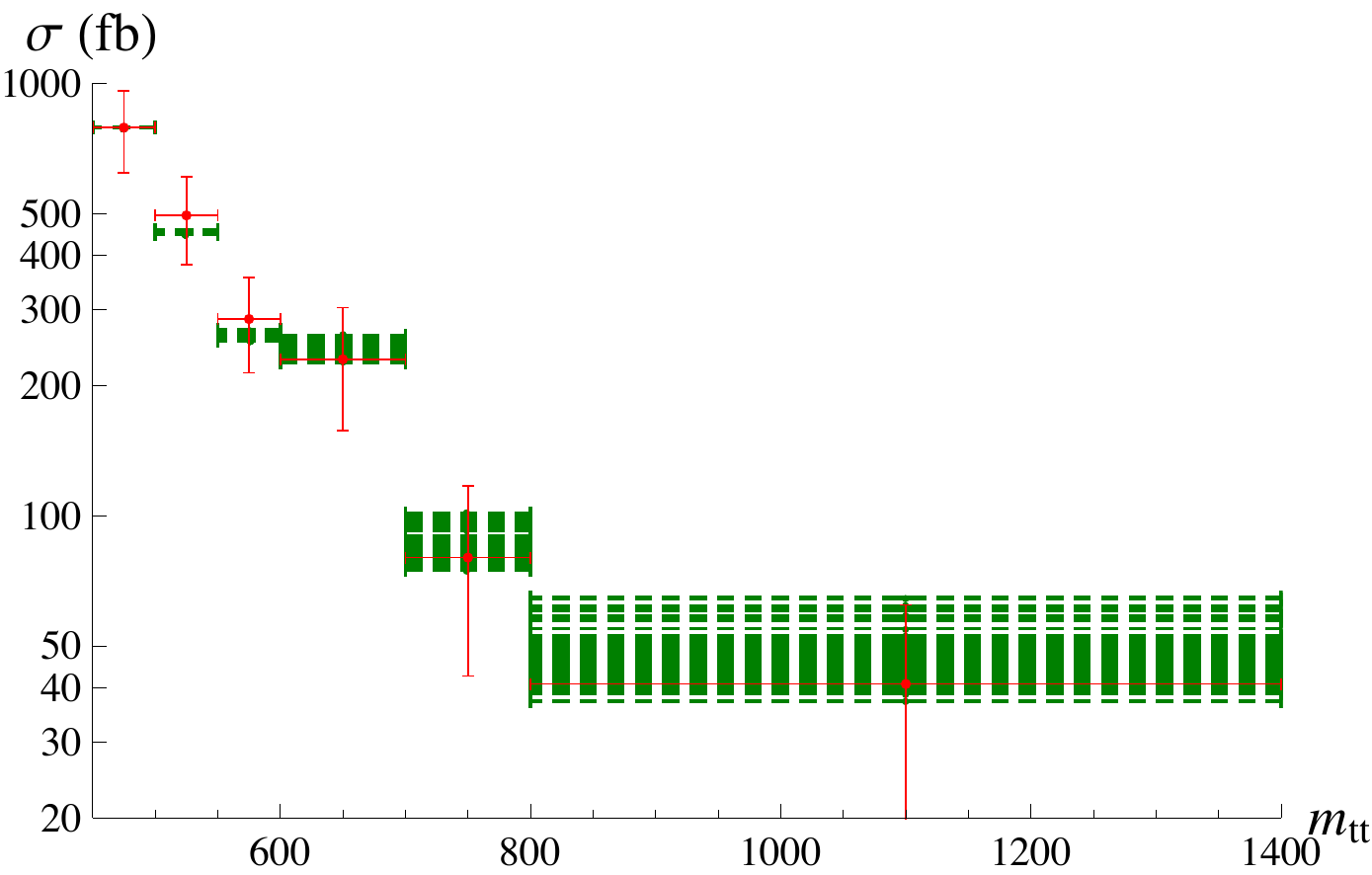}
\caption{The $t\bar t$ production cross-section as a function of the 
$t\bar t$ invariant mass $m_{tt}$, for $M_G=1.5$~TeV.  The crosses are
the experimental values. The band represent the solutions of the
scattered plot in Figure~\ref{scat15}  that are within 1.5 $\sigma$ of
the data. 
}
\label{mttdist_15}
\end{figure}
\begin{figure}
\includegraphics[scale=0.60]{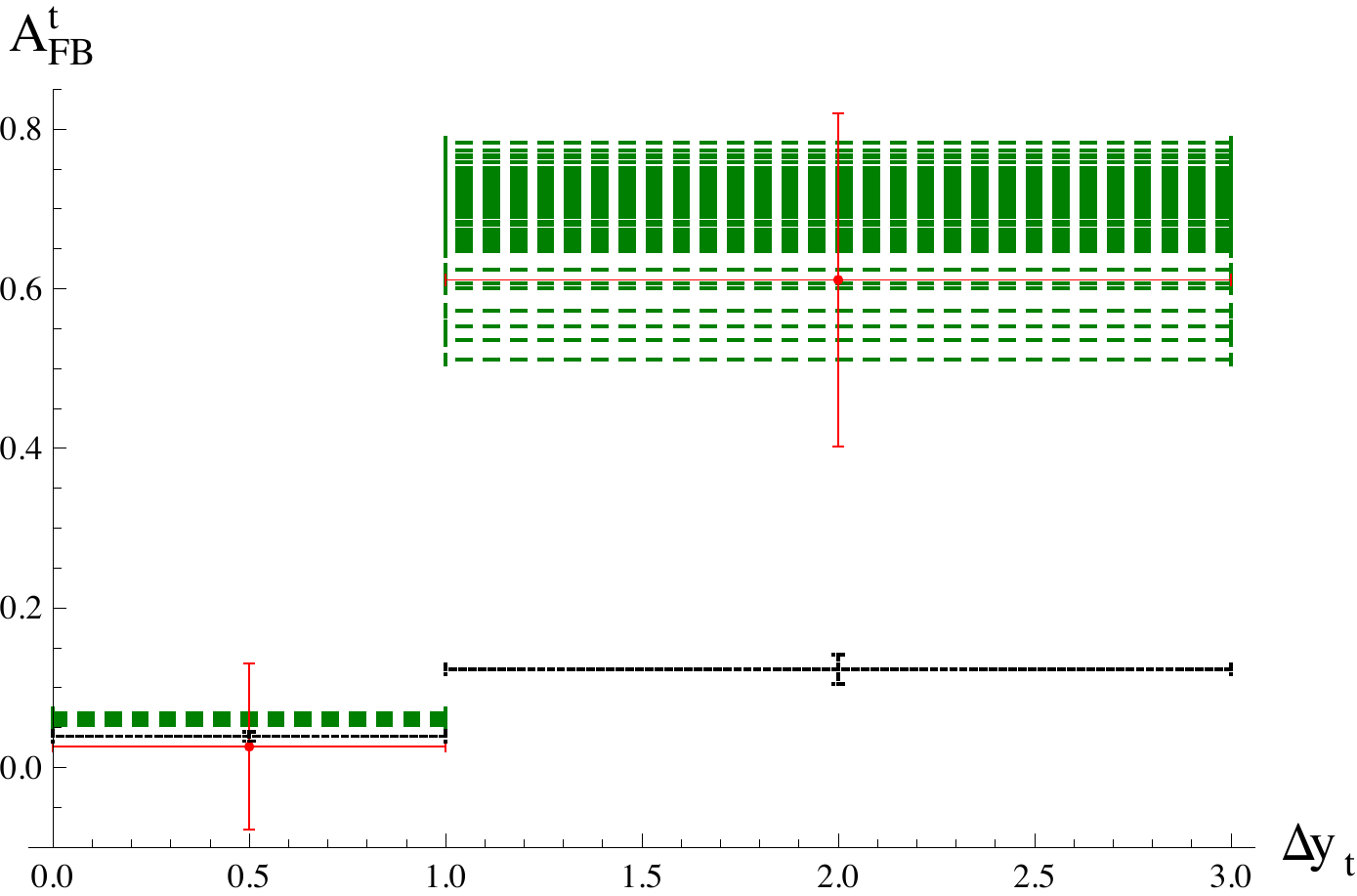}
\caption{The rapidity dependence of the forward-backward asymmetry in
  $t\bar t$ production. for $M_G=1.5$~TeV.  
 The crosses are
the experimental values. The band represent the solutions of the
scattered plot in Figure~\ref{scat15}  that are within 1.5 $\sigma$ of
the data. 
}
\label{afbvsdy_15}
\end{figure}

To summarize, in this section we have limited the parameter space of the
effective theory defined in the previous section, with the requirement
that it gives a significant deviation in $A_{FB}^t$ while not
violating any known bounds, including  those from flavor physics and
all Tevatron top quark data. 
This bottom-up approach allows us to reduce the parameter space of the
effective theory to the point of having a rather specific description
of the new interactions of known quarks.  We conclude that it is
possible to generate a significant deviation from the SM in $A_{FB}^t$
as observed at the Tevatron, while respecting all existing bounds,
including the flavor constraints. This is not in contradiction with
the results of Ref.~\cite{sekhar}, since in that case the axi-gluon  model 
used (the same as in Ref.~\cite{frampton}), is more constrained by
flavor physics since in it the freedom to have universal couplings and
still have a significant deviation in $A_{FB}^t$ is absent due to the
necessary choices of the quark couplings. 

In the next section,  we consider the possibility that quarks of a fourth
generation are strongly coupled to the new interaction. We will use
the remaining  parameter space of the theory, with the color-octet
mass, width and  couplings  to light quarks constrained, in order to predict the
production of fourth-generation quarks at the Tevatron as well as at
the LHC.

\vspace*{0.3cm}
\section{A Heavy Fourth Generation} 
\label{fg}
Having limited the parameter space of our effective theory by physics
observables related to the first three generations, in this Section we use
the resulting model to make predictions for the production of
fourth-generation quarks both at the Tevatron and at the LHC. 

As mentioned in Section~\ref{intro}, the presence of fourth-generation
quarks is motivated as an alternative mechanism of EWSB.
 For the new strong interaction to be at around the
TeV scale, a natural choice, the dynamical masses of condensing
fermions should be close to $(500-600)~$GeV~\cite{bhl}. 
Current bounds on the masses of the fourth-generation quarks from
direct searches are $m_{U_4}> 335~$GeV~\cite{u4bound}, and $m_{D_4}>
385~$GeV~\cite{d4bound}, both at $95\%$~C.L., below the condensation
model values, but not too far from them. 
We focus here on the most
constraining bounds, which come from electroweak precision
measurements. In Ref.~\cite{graham} it is shown that it is possible to
accommodate a heavy fourth generation with some restrictions on the
mass spectrum, as well as the Higgs mass. 
Here, we will not assume anything about $m_h$, although the type of
theories  that result in these phenomenological models have typically
larger $m_h$ values than in SM fits, which are  compatible with the
findings of Ref.~ \cite{graham}.

The most constrained parameter of a fourth-generation quark sector
with this typical mass scale is the mass difference, since it affects
the $T$ parameter giving a positive contribution to it.  Having a positive contribution   
to $T$ is actually good since it allows for  larger values of $S$,
which is also greatly constrained by electroweak measurements. 
A degenerate fourth family of quarks contributes with $\delta
S_q\simeq 0.2$. The $S-T$ fits allow typically for larger values of
$S$ and $T$ as long as $S\sim T$. On the other hand, the fits disfavor
values of $T$ much larger than  $0.3$, even if $m_h$ is heavy. This
translates in  the approximate bound~\cite{graham} $|m_{U_4}-m_{D_4} |<
M_W$. Somewhat larger values are still compatible with
electroweak fits in some region of the parameters. However,  we 
will consider mass differences below $M_W$
in order to limit the amount of $T$ from this one source. This choice
has important phenomenological consequences since it suppresses 
intergenerational weak transitions such as $D_4\to U_4$ by requiring a
3-body phase space. Thus, if the mass difference is significantly
below $M_W$ the 2-body weak decays to third generation quarks will be
favored as long as the CKM mixing between the third and fourth
generation quarks is not too small.

\vspace*{0.3cm}
\section{Predictions for the Tevatron and the LHC}
\label{predictions}

Here we  consider the reach of the Tevatron to observe the
pair-production of  fourth-generation  quarks with masses of 
$m_{4}=500~$GeV, via the
interactions described by (\ref{leff}). The choice of this value is
mainly for concreteness, but at the same time is  motivated by several
arguments.
As previously mentioned, in models of
EWSB via fourth-generation condensation~\cite{bhl,bd} the dynamically
generated fermion mass is typically in the range $\simeq (500-600)~$GeV when the scale of the
condensing interaction is $O(1)~$TeV. Then, if the effective theory
described in Section~\ref{efftheory} is to emerge from this scenario,
these are the typical values of the four-generation quark masses.
Finally and as we will see below, the Tevatron is already beginning to
be sensitive to these effective theories for quark masses around this
range. Thus, although this specific value is just  a straw-man choice, the
actual value of the heavy quark masses should not be further than
$O(100)~$GeV from it. 

\begin{figure}
\includegraphics[scale=0.75]{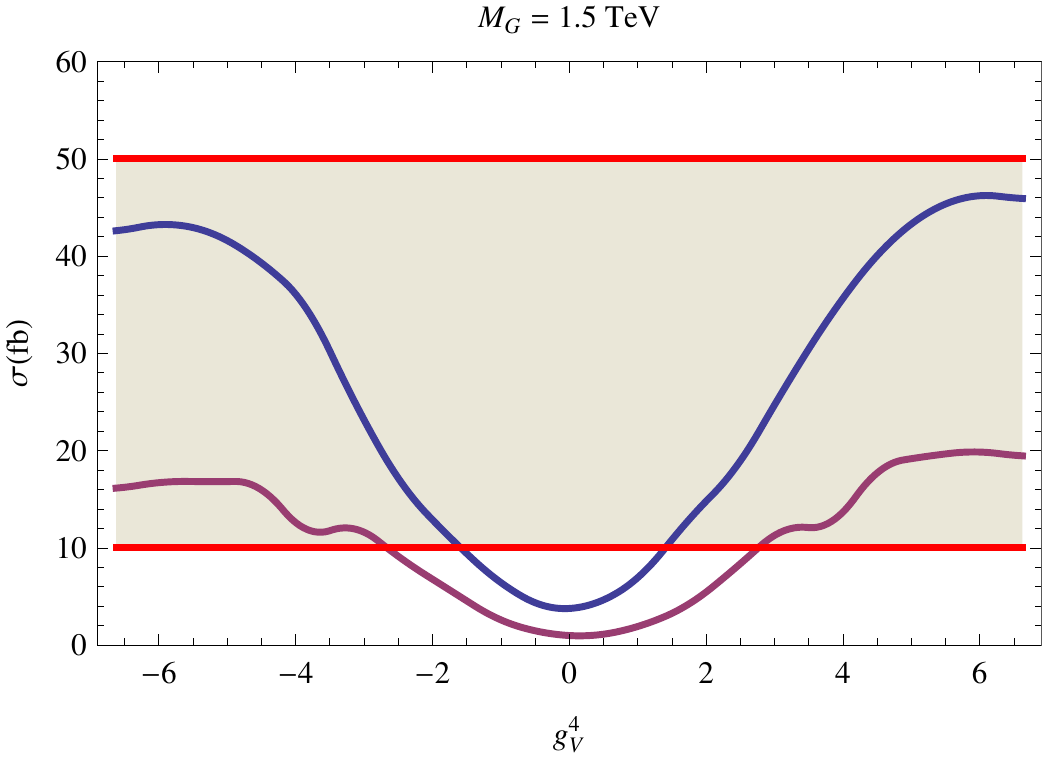}
\caption{$U_4$ pair-production cross section vs. the $U_4$ vector
 coupling $g^4_V$ for $g^4_A=0$, and  for $M_G=1.5$~TeV. }
\label{u415}
\end{figure}

We first consider the $U_4\bar U_4$ pair-production cross section at
the Tevatron. For the fixed values of $m_{U_4}$ we are considering
here, we will use values of the $G'$ couplings to light quarks that 
are consistent with all bounds plus give a significant increase in
$A_{FB}^t$. These values depend on  $M_{G'}$, and can be obtained from
the solutions displayed in Figures~\ref{afbgtr1} and \ref{afbgtr15}.  
Finally, we need to  fix the $G'$ coupling to $U_4$. Since a large
fraction of the $U_4$ production cross-section comes from the
interference with the SM s-channel gluon, and this only depends on the
vector coupling $g^4_V = (g^4_L + g^4_R)/2$, we will fix the axial
coupling $g^4_A=0$ just for concreteness, and we will study the
dependence of the cross-section with $g^4_V$. 
\begin{figure}
\includegraphics[scale=0.75]{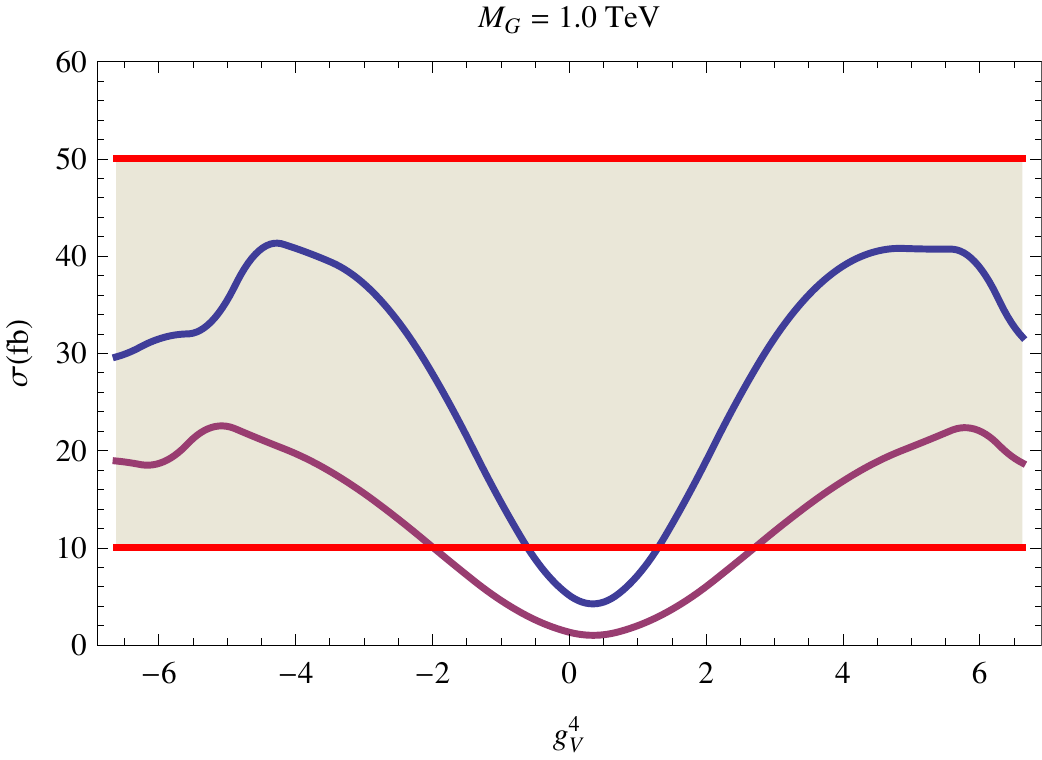}
\caption{$U_4$ pair-production cross section vs. the $U_4$ vector
 coupling $g^4_V$ for $g^4_A=0$, and  for $M_G=1$~TeV. }
\label{u41}
\end{figure}
In Figure~\ref{u41} we plot the production cross section of $U_4$
pairs from  $p\bar p$ collisions at  $\sqrt{s}=1.96~$TeV, as a function of the $U_4$ vector coupling
$g^4_V$ for a color-octet mass of $M_G=1~$TeV. The top line corresponds to $m_{U_4}=450$~GeV, whereas the
bottom one is for $m_{U_4}=500~$GeV.
The top horizontal line is an approximate value of the Tevatron
sensitivity for $5~fb^{-1}$, whereas the bottom horizontal line is
intended as an estimate of the future reach, assuming the two-body
channel $U_4\to b W^+$ dominates.
We can see that the
Tevatron will have reach  to observe a strongly-coupled fourth-generation up-type quark
with couplings large enough  to trigger condensation and EWSB. The
plots reflect a particular solution for the couplings of light quarks
to the color-octet that gives a large increase in the
$A_{FB}^t$ while  passing all constraints. Similarly, in
Figure~\ref{u415} we show the results for $M_G=1.5~$TeV, where the
solution for the light quark couplings to $G'$ is now different than
in the previous example. 
As we can see, despite the increase in the color-octet mass, the
Tevatron reach is still significant. This is due to the requirement 
that the light-quark couplings be large enough to give a large
deviation in $A_{FB}^t$, which means that they have to be larger than
for lighter $G'$ masses.  

Analogously, we can consider the production of the down-type quark
$D_4$ at the Tevatron with similar results. In models where only one
of the two fourth-generation quarks condenses, the production of the
non-condensing one is somewhat smaller than that of the condensing
quark, since its couplings to the color-octet interaction are
smaller.  Thus, if for instance only the $U_4$ were to condense to
break the electroweak symmetry, the $D_4$ production cross section could be well below
the Tevatron sensitivity. On the other hand, if both are
super-critically coupled to the color-octet, they will have very
similar production cross sections. 

\begin{figure}[t]
\includegraphics[scale=0.75]{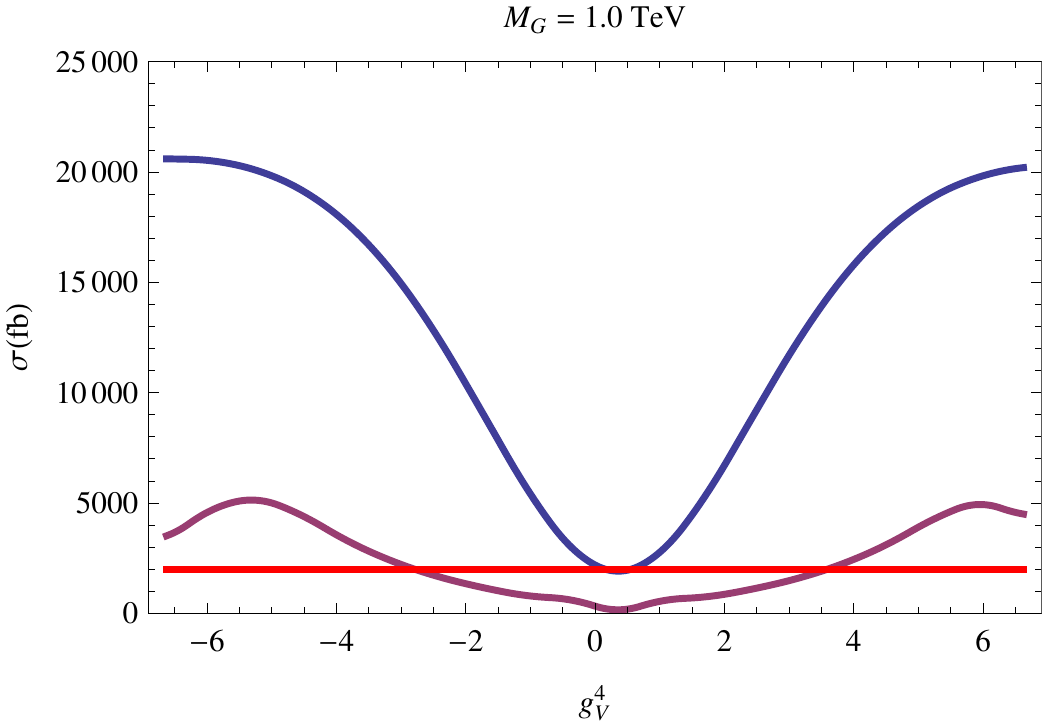}
\caption{$U_4$ pair-production cross section vs. the $U_4$ vector
 coupling $g^4_V$ for $g^4_A=0$, and  for $M_G=1$~TeV, for
 $\sqrt{s}=7~$TeV. The top line is for $m_{U_4} =450~$GeV, the second
 one for $m_{U_4} =600~$GeV, whereas the horizontal line corresponds
 to the SM QCD production.}
\label{u4lhc_1}
\end{figure}

\begin{figure}
\includegraphics[scale=0.75]{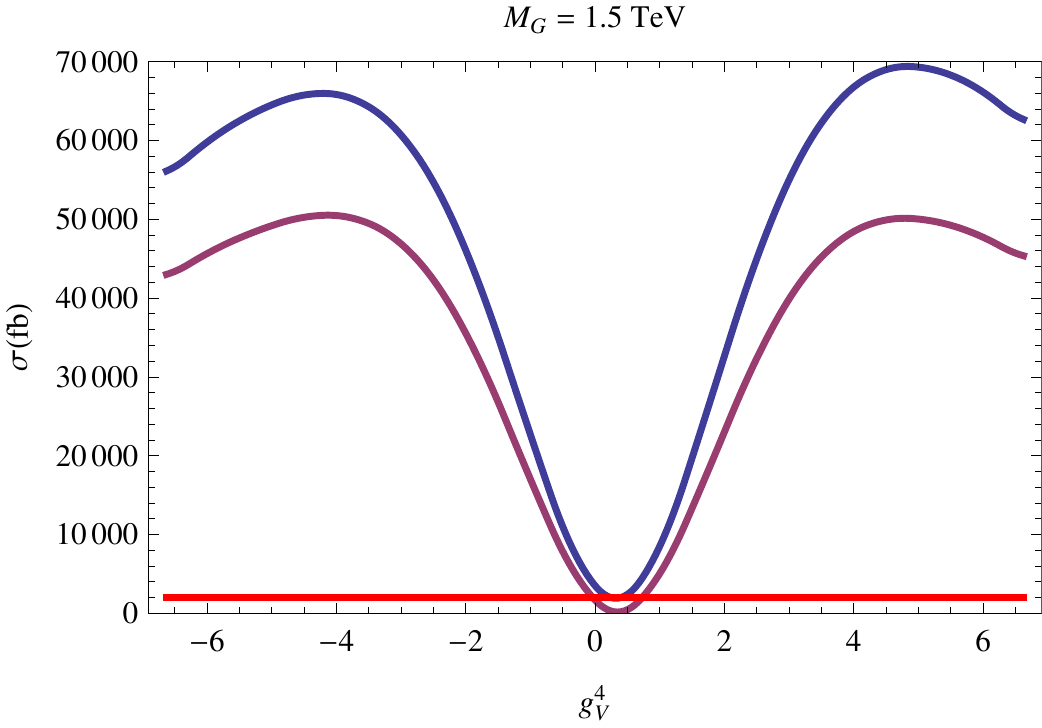}
\caption{$U_4$ pair-production cross section vs. the $U_4$ vector
 coupling $g^4_V$ for $g^4_A=0$, and  for $M_G=1.5$~TeV, for
 $\sqrt{s}=7~$TeV. The top line is for $m_{U_4} =450~$GeV, the second
 one for $m_{U_4} =600~$GeV, whereas the horizontal line corresponds
 to the SM QCD production.}
\label{u4lhc_15}
\end{figure}

We finally address the implications for the LHC. 
We have significantly reduced the parameter space of the effective
theory  described by (\ref{leff}).  by imposing light flavor universality, 
asking for a deviation in $A_{FB}^t$, and for strong enough couplings for
fourth-generation condensation while respecting limits from Tevatron. 
We can now make predictions for the
LHC using this well-defined region of 
parameter space.
Unlike in the AdS$_5$ model of Ref.~\cite{bd}, where QCD completely
dominates the pair-production of fourth-generation quarks, 
in this case it will be
dominated by the color-octet contribution.
This is due in part to the relatively light masses we are considering
here ($1-1.5$~TeV), but also to the strong couplings necessary to make
condensation viable. 

To compare to the Tevatron results above, we consider the LHC with $7$
~TeV center-of-mass energy.  For illustration, we use the values of the light quark
couplings to the color-octet $G'$ that result in an excess in
$A_{FB}^t$, as shown above. Once again, we use $g_A^4=0$ for
concreteness, but the heavy-quark production cross section does not
depend on this coupling significantly. The results for $U_4$
pair-production are shown in Figures~\ref{u4lhc_1} and \ref{u4lhc_15},
for $M_G = 1~$TeV and $M_G=1.5~$TeV, respectively.
We see that the parameter space selected from the effective theory
presented in Section~\ref{efftheory} in order to give an
excess in $A_{FB}^t$ consistent with observations, can potentially
result in very large signals in the pair-production of
fourth-generation quarks mediated by the new color-octet interaction. 
This, despite the fact that this very same set of parameters has not
yet been excluded by the Tevatron. The LHC running at
$\sqrt{s}=7~$TeV, and accumulating $1~$fb$^{-1}$ should be able to
exclude a large fraction of the relevant parameter space, as it is clearly seen
in Figures~\ref{u4lhc_1} and \ref{u4lhc_15}.

\section{Summary and Conclusions}
\label{conclusions}
We have considered a scenario where the 
 new physics at the TeV scale responsible for EWSB is  coupled to
 flavor. In particular, if the new interaction is strongly coupled to heavier
 generations, such as the third and/or a hypothetical fourth
 generation, the phenomenology of this flavor dependence will be very
 distinct. We have taken an effective theory approach to this problem,
 by adding to the SM just the minimum ingredients needed for this
 scenario: a color-octet massive state, and a fourth generation
 strongly coupled to it, which presumable will lead to EWSB through
 the condensation of at least one of its quarks. 
 We have shown that if we require  a significant contribution to
 $A_{FB}^t$, 
 and we impose  the constraints from flavor physics, the parameter space
 of the effective theory is greatly reduced, making it quite
 predictive.   This can be seen in the progression from
 Figure~\ref{scat1} to \ref{afbvsdy_15}.

Adding the requirement that at least one of the fourth-generation
quarks is strongly coupled to the new interaction, results in
predictions for the Tevatron that could be falsified before the end of
Run~II.  These predictions are summarized in Figures~\ref{u41} and
\ref{u415}. 
 Furthermore, we predict that this scenario can be easily
observed/excluded at the LHC with $\sqrt{s}=7$~TeV and $1~{\rm fb}^{-1}$
of integrated luminosity, as it can be seen in Figures~\ref{u4lhc_1}
and \ref{u4lhc_15}. 

The bottom-up approach used here is complementary to model-building. 
If the deviation in $A_{FB}^t$ is confirmed, and the LHC observes the
production of fourth-generation quarks with cross-sections similar to
those shown in the previous section, it would be evidence that the new
interaction is coupled to flavor and that the new heavy quarks have
indeed a role in EWSB.  Coupled to the flavor constraints, we are left
with an effective theory  where the only fermion of the first three
generations strongly coupled to the new interaction is the
right-handed top quark $t_R$. Building such theory, although
challenging, would be a step towards understanding EWSB and flavor.

\noindent{\bf Acknowledgments } 
The authors  acknowledge the support of the State of S\~{a}o Paulo
Research Foundation (FAPESP).
G.~B. also acknowledges the support of the John Simon Guggenheim
Foundation and  Brazilian  National Council
for Technological and Scientific Development (CNPq), and thanks the Fermilab
Theory Group, where this work was started, for its hospitality.

\end{document}